
\input epsf.tex
\magnification=\magstep0
\hsize=13.5 cm               
\vsize=19.0 cm               
\baselineskip=12 pt plus 1 pt minus 1 pt  
\parindent=0.5 cm  
\hoffset=1.3 cm      
\voffset=2.5 cm      
\font\twelvebf=cmbx10 at 12truept 
\font\twelverm=cmr10 at 12truept 
\overfullrule=0pt
\nopagenumbers    
%
\newtoks\leftheadline \leftheadline={\hfill {\eightit Authors' name}
\hfill}
\newtoks\rightheadline \rightheadline={\hfill {\eightit the running title}
 \hfill}
\newtoks\firstheadline \firstheadline={{\eightrm Bull. Astron. Soc.
India (1998) {\eightbf xx,} } \hfill}
\def\makeheadline{\vbox to 0pt{\vskip -22.5pt
\line{\vbox to 8.5 pt{}\ifnum\pageno=1\the\firstheadline\else%
\ifodd\pageno\the\rightheadline\else%
\the\leftheadline\fi\fi}\vss}\nointerlineskip}
%
\font\eightrm=cmr8  \font\eighti=cmmi8  \font\eightsy=cmsy8
\font\eightbf=cmbx8 \font\eighttt=cmtt8 \font\eightit=cmti8
\font\eightsl=cmsl8
\font\sixrm=cmr6    \font\sixi=cmmi6    \font\sixsy=cmsy6
\font\sixbf=cmbx6
%
\def\eightpoint{\def\rm{\fam0\eightrm}
\textfont0=\eightrm \scriptfont0=\sixrm \scriptscriptfont0=\fiverm
\textfont1=\eighti  \scriptfont1=\sixi  \scriptscriptfont1=\fivei
\textfont2=\eightsy \scriptfont2=\sixsy \scriptscriptfont2=\fivesy
\textfont3=\tenex   \scriptfont3=\tenex \scriptscriptfont3=\tenex
\textfont\itfam=\eightit  \def\it{\fam\itfam\eightit}%
\textfont\slfam=\eightsl  \def\sl{\fam\slfam\eightsl}%
\textfont\ttfam=\eighttt  \def\tt{\fam\ttfam\eighttt}%
\textfont\bffam=\eightbf  \scriptfont\bffam=\sixbf
\scriptscriptfont\bffam=\fivebf \def\bf{\fam\bffam\eightbf}%
\normalbaselineskip=10pt plus 0.1 pt minus 0.1 pt
\normalbaselines
\abovedisplayskip=10pt plus 2.4pt minus 7pt
\belowdisplayskip=10pt plus 2.4pt minus 7pt
\belowdisplayshortskip=5.6pt plus 2.4pt minus 3.2pt \rm}
%
%
\def\leftdisplay#1\eqno#2$${\line{\indent\indent\indent%
$\displaystyle{#1}$\hfil #2}$$}
\everydisplay{\leftdisplay}
%
\def\frac#1#2{{#1\over#2}}


%
%
\def\pmb#1{\setbox0=\hbox{$#1$}\kern-0.015em\copy0\kern-\wd0%
\kern0.03em\copy0\kern-\wd0\kern-0.015em\raise0.03em\box0}
%
\pageno=1
\vglue 60 pt  
%
\leftline{\twelvebf  Physics Input from Multiwavelength Observations of AGN} 
%
\smallskip
\vskip 46 pt  
\leftline{\twelverm M. B\"ottcher\footnote{$^1$}{\tenrm Chandra Fellow}} 
\vskip 4 pt
\leftline{\eightit Physics and Astronomy Department, Rice University, MS 108,
Houston, TX 77005-1892, USA}
%
%
\vskip 20 pt 
%
%
\leftheadline={\hfill {\eightit M. B\"ottcher} \hfill}
\rightheadline={\hfill {\eightit Multiwavelength Modeling of AGN 
}  \hfill}

%
{\parindent=0cm\leftskip=1.5 cm

{\bf Abstract.}
\noindent 
The current status of leptonic jet models for blazars 
is reviewed. Differences between the quasar and BL-Lac 
subclasses of blazars may be understood in terms of 
the dominance of different radiation mechanisms in 
the gamma-ray regime. Spectral variability patterns of 
different blazar subclasses appear to be significantly 
different and require different intrinsic mechanisms 
causing gamma-ray flares. As examples, recent results of 
long-term multiwavelength monitoring of PKS 0528+134, 
3C 279, and Mrk 501 are interpreted in the framework 
of leptonic jet models. Short-term variability patterns
give important additional clues about the source geometry
and the relevant radiation mechanisms in blazars. Challenges 
for future observational efforts are discussed. 
\smallskip 
\vskip 0.5 cm  
{\it Key words:} Active galaxies, extragalactic jets, theory,
radiation mechanisms
}                                 
%
%
%

\vskip 20 pt
\centerline{\bf 1. Introduction}
\bigskip
\noindent 
Recent high-energy detections and simultaneous broadband 
observations of blazars, determining their spectra and spectral 
variability, are posing strong constraints on models of blazars. 
66 blazars have been detected by EGRET at energies above 100~MeV 
(Hartman et al. 1999), the two nearby high-frequency peaked 
BL~Lac objects (HBLs) Mrk~421 and Mrk~501 are now multiply 
confirmed sources of multi-GeV -- TeV radiation (Punch et al.
1992, Petry et al. 1996, Quinn et al. 1996, Bradbury et al.
1997), and the TeV detections of PKS~2155-314 (Chadwick et
al. 1999) and 1ES~2344+514 (Catanese et al. 1998) are awaiting 
confirmation. Most EGRET-detected blazars exhibit rapid variability
(e.g., Mukherjee et al. 1997), in some cases on intraday and even 
sub-hour (e. g., Gaidos et al. 1996) timescales, where generally 
the most rapid variations are observed at the highest photon 
frequencies.

The broadband spectra of blazars consist of at least two 
clearly distinct spectral components (e.g., von Montigny et
al. 1995). The first one extends in the case of flat-spectrum 
radio quasars (FSRQs) from radio to optical/UV frequencies, 
in the case of HBLs up to soft and even hard X-rays, and is 
consistent with non-thermal synchrotron radiation from 
ultrarelativistic electrons. The second spectral component 
emerges at $\gamma$-ray energies and peaks at several
MeV -- a few GeV in most quasars, while in the case of some 
HBLs the peak of this component appears to be located at TeV 
energies. 

\midinsert
\epsfysize=6cm
\hskip 3cm \epsffile{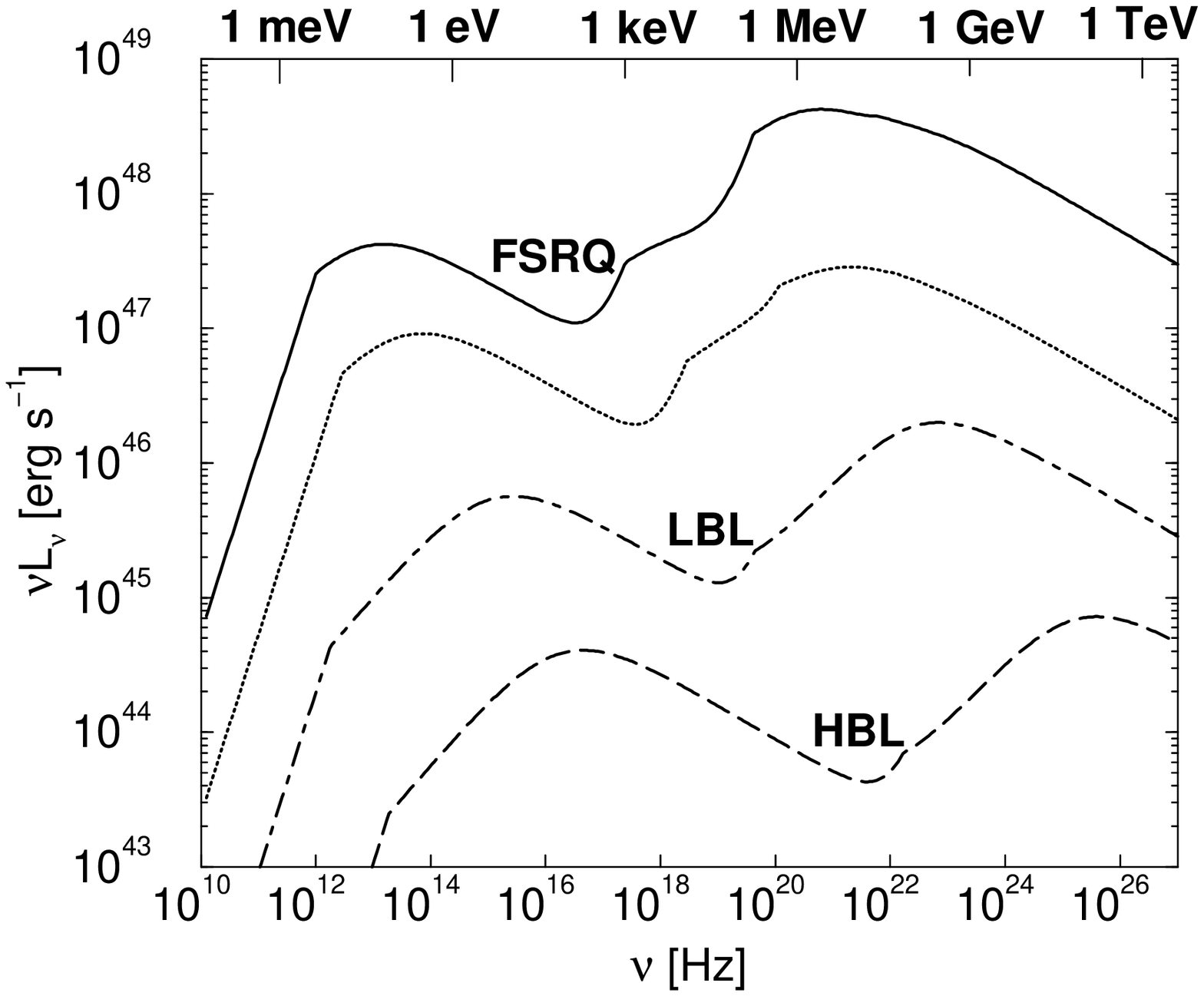}
\medskip
\noindent
{\bf Figure 1.} Schematic, prototypical broadband spectra of
different blazar classes.
\endinsert

The different sub-classes of blazars exhibit markedly different
spectral variability patterns. In quasars, the relative amplitude
of a $\gamma$-ray flare, $\delta L_{\rm HE}$, is generally larger 
than the corresponding variability amplitude at lower energies 
(i.e. in the synchrotron component), $\delta L_{\rm LE}$, and the
$\gamma$-ray dominance during flares may be super-quadratic, i.e.
$\delta L_{\rm HE} \sim \delta L_{\rm LE}^{\alpha}$ with $\alpha
\ge 2$ (e.g., Wehrle et al. 1998). The $\gamma$-ray specta generally
harden during $\gamma$-ray flares, although the $\nu F_{\nu}$ peak
of either of the spectral components has so far not been seen to 
shift in any systematic way (see, however, B\"ottcher [1999] for a
prediction on the expected synchrotron peak shift). Detailed
short-term variability studies show generally no consistent
patterns of timelags between different energy bands (e.g.,
Hartman et al. 2001b).

Spectral variability of high-frequency peaked BL Lac objects is
also generally characterized by a higher variability amplitude in
$\gamma$-rays than at lower frequencies, but here we generally
find a sub-quadratic amplitude behavior, i. e. $\delta L_{\rm HE} 
\sim \delta L_{\rm LE}^{\alpha}$ with $\alpha \le 2$ (e.g.,
Petry et al. 2000). HBL flares are accompanied by spectral
hardening in both components, and the synchrotron peak shifts
significantly to higher frequency during flares (e.g., Takahashi
et al. 1996, Pian et al. 1998, Urry et al. 1997). Within the
synchrotron emission component, there is a consistent time-lag 
trend with flares at higher frequencies systematically preceding 
lower-frequency flares by an amount consistent with being due to
synchrotron cooling (e.g., Takahashi et al. 1996, Edelson et al.
1995). More detailed analyses of time-resolved HBL spectra during
flaring activity often show a clockwise rotation when plotted in
a hardness-intensity diagram (e.g., Takahashi et al. 1996). 

The bolometric luminosity of EGRET-detected quasars and some
low-frequency peaked BL~Lac objects (LBLs) during flares is
dominated by the $\gamma$-ray emission. If this emission were 
isotropic, it would correspond to enormous luminosities (up to 
$\sim 10^{49}$~erg~s$^{-1}$) which, in combination with the
short observed timescales (implying a small size of the 
emission region) would lead to a strong modification of the
emissivity spectra by $\gamma\gamma$ absorption, in contradiction 
to the observed smooth power-laws at EGRET energies. This has 
motivated the concept of relativistic beaming of radiation 
emitted by ultrarelativistic particles moving at relativistic 
bulk speed along a jet (for a review of these arguments, see 
Schlickeiser 1996). While it is generally accepted that 
blazar emission originates in relativistic jets, the radiation 
mechanisms responsible for the observed $\gamma$ radiation are 
still under debate. It is not clear yet whether in these jets 
protons are the primarily accelerated particles, which then 
produce the $\gamma$ radiation via photo-pair and photo-pion 
production, followed by $\pi^0$ decay and synchrotron emission 
by secondary particles (e. g., Mannheim 1993) or by the primarily
accelerated protons themselves (Aharonian 2000, M\"ucke \& 
Protheroe 2000), or electrons (and positrons) are accelerated 
directly and produce $\gamma$-rays in Compton scattering 
interactions with the various target photon fields in the 
jet (Marscher \& Gear 1985, Maraschi et al. 1992, Dermer et 
al. 1992, Sikora et al. 1994, Ghisellini \& Madau 1996,
B\"ottcher et al. 1997, Bla$\dot{\rm z}$ejowski et al. 2000). 

In this review, I will describe the current status of blazar
models based on leptons as the primary constituents of the 
jet which are responsible for the $\gamma$-ray emission. For
a recent review of hadronic jet models see, e.g., Rachen (1999).
In Section 2, I will give a description of the common features
present in the different variations of these models and discuss 
the different $\gamma$-ray production mechanisms and their 
relevance for different blazar classes. In Section 3, I will 
review recent progress in understanding intrinsic differences 
between different blazar classes. In Section 4, I will discuss 
broadband spectral variability of individual blazars and their 
interpretation in the framework of leptonic jet models. Finally,
in Section 5, I will discuss some open questions with respect
to our understanding of blazar spectra and variability, and
challenges for future observations to address some of these 
questions.

\vskip 20 pt
\centerline{\bf 2. Salient features of leptonic jet models}
\bigskip 
\noindent
The basic geometry of leptonic blazar jet models is illustrated
in Fig. 2. At the center of the AGN, an accretion disk around a 
supermassive, probably rotating, black hole is powering a 
relativistic jet. Along this pre-existing jet structure,
occasionally blobs of ultrarelativistic electrons are ejected 
at relativistic bulk velocity. 

\midinsert
\epsfysize=6cm
\hskip 3cm \epsffile{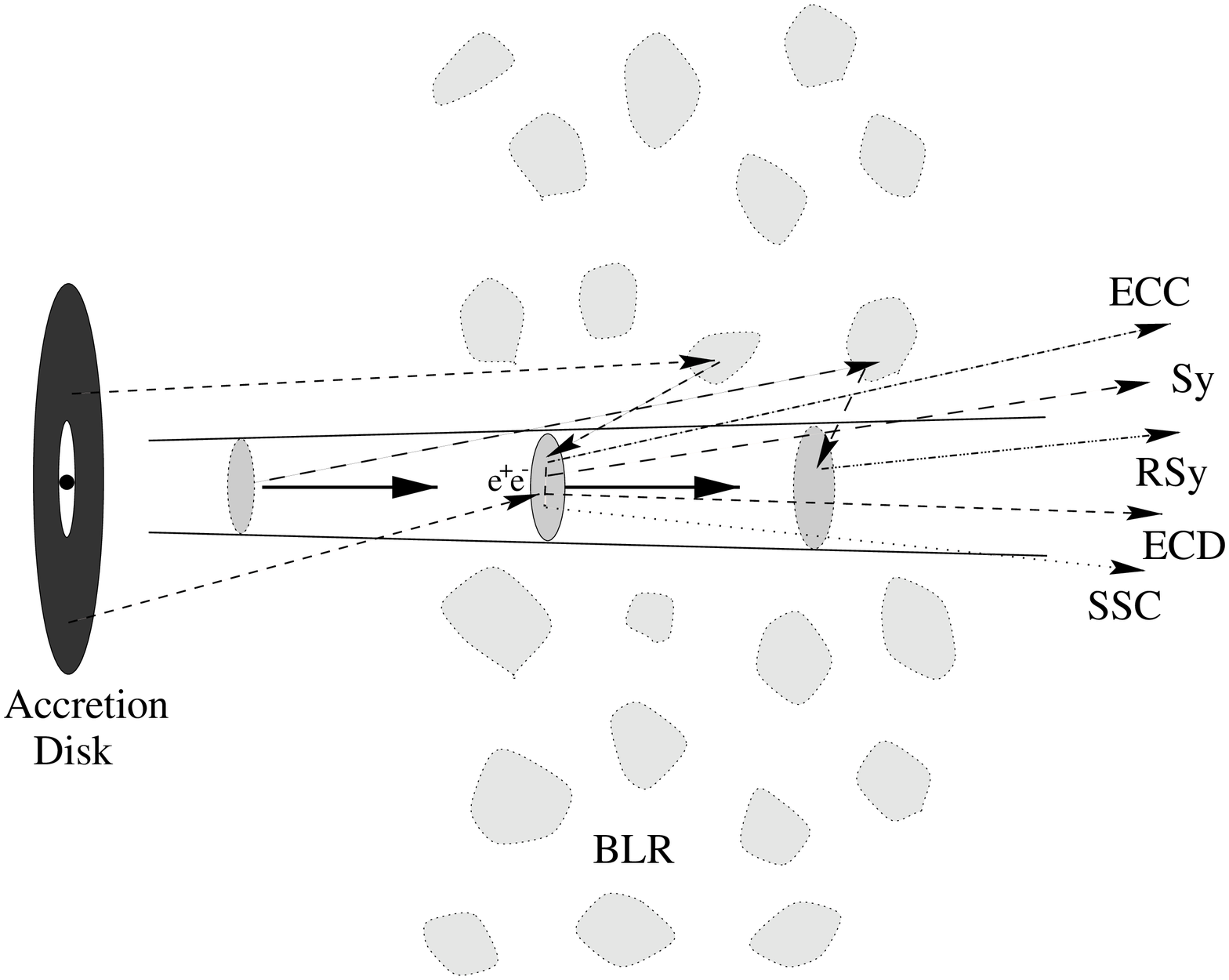}
\medskip
\noindent
{\bf Figure 2.} Illustration of the model geometry and the relevant 
$\gamma$ radiation mechanisms for leptonic jet models.
\endinsert

The electrons are emitting synchrotron radiation, which will 
be observable at IR -- UV or even X-ray frequencies, and hard 
X-rays and $\gamma$-rays via Compton scattering processes. Possible 
target photon fields for Compton scattering are the synchrotron photons
produced within the jet (the SSC process, Marscher \& Gear 1985,
Maraschi et al. 1992, Bloom \& Marscher 1996, Georganopoulos \&
Marscher 1998b), the UV -- soft X-ray emission from the disk 
--- either entering the jet directly (the ECD [External 
Comptonization of Direct disk radiation] process; Dermer 
et al. 1992, Dermer \& Schlickeiser 1993) or after reprocessing 
at the broad line regions or other circumnuclear material (the 
ECC [External Comptonization of radiation from Clouds] process; 
Sikora et al. 1994, Blandford \& Levinson 1995, Dermer et al.
1997), or jet synchrotron radiation reflected at the broad line 
regions (the RSy [Reflected Synchrotron] mechanism; Ghisellini
\& Madau 1996, Bednarek 1998, B\"ottcher \& Dermer 1998). It has
recently also been suggested that infrared emission from circumnuclear
dust may play a significant role as seed photon field for Compton
scattering in the jet (Bla$\dot{\rm z}$ejowski et al. 2000, 
Abeiter et al. 2001).

The relative importance of these components may be estimated by
comparing the energy densities of the respective target photon
fields. Denoting by $u'_B$ the co-moving energy density of the 
magnetic field, the energy density of the synchrotron radiation
field, governing the luminosity of the SSC component, may be 
estimated by $u'_{sy} \approx u'_B \, \tau_T \, \gamma_e^2$,
where $\tau_T = n'_{e, B} \, R'_B \, \sigma_T$ is the Thomson depth 
of the relativistic plasma blob and $\gamma_e$ is the average
Lorentz factor of electrons in the blob. The SSC spectrum exhibits
a broad hump without strong spectral break, peaking around 
$\langle\epsilon\rangle_{SSC} \approx (B' / B_{cr}) \, D \,
\gamma_e^4 \approx \langle\epsilon\rangle_{sy} \, \gamma_e^2$, 
where $B'$ is the co-moving magnetic field, $B_{cr} = 4.414 
\cdot 10^{13}$~G, and $D = \left( \Gamma \, [1 - \beta_{\Gamma}
\cos\theta_{obs}] \right)^{-1}$ is the Doppler factor associated
with the bulk motion of the blob. Throughout this paper, all 
photon energies are described by the dimensionless quantity 
$\epsilon = h \nu / (m_e c^2)$. 

If the blob is sufficiently far from the central engine of the AGN
so that the accretion disk can be approximated as a point source of
photons, its photon energy density (in the co-moving frame) is 
$u'_D \approx L_D / (4 \pi \, z^2 \, c \, \Gamma^2)$, where $L_D$
is the accretion disk luminosity, and $z$ is the height of the blob
above the accretion disk. The ECD spectrum can exhibit a strong
spectral break, depending on the existence of a low-energy cutoff
in the electron distribution function, and peaks at $\langle\epsilon
\rangle_{ECD} \approx \langle\epsilon\rangle_D \, (D / \Gamma) \,
\gamma_e^2$, where $\langle\epsilon\rangle_D$ is the average photon
energy of the accretion disk radiation (typically of order $10^{-5}$
for geometrically thin, optically thick accretion disks around black
holes of $\sim 10^8$ -- $10^{10} \, M_{\odot}$).

Part of the accretion disk and the synchrotron radiation will be
reprocessed by circumnuclear material in the broad line region
and can re-enter the jet. Since this reprocessed radiation is 
nearly isotropic in the rest-frame of the AGN, it will be strongly
blue-shifted into the rest-frame of the relativistically moving
plasma blob. Thus, assuming that a fraction $a_{BLR}$ of the radiation
is rescattered into the jet trajectory, we find for the energy density 
of rescattered accretion disk photons: $u'_{ECC} \approx L_D \, a_{BLR}
\, \Gamma^2 / (4 \pi \, \langle r \rangle_{BLR}^2 \, c)$, where
$\langle r \rangle_{BLR}$ is the average distance of the BLR material
from the central black hole. The ECC photon spectrum peaks around
$\langle\epsilon\rangle_{ECC} \approx \langle\epsilon\rangle_D \,
D \, \Gamma \, \gamma_e^2 \approx \langle\epsilon\rangle_{ECD} \, 
\Gamma^2$. 

For the synchrotron mirror mechanism, additional constraints due to
light travel time effects need to be taken into account in order to
estimate the reflected synchrotron photon energy density (for a
detailed discussion see B\"ottcher \& Dermer 1998), which is well 
approximated by $u'_{RSy} \approx u'_{sy} \, 4 \,\Gamma^3 \, a_{BLR} 
\, (R'_B / \Delta r_{BLR}) \, ( 1 - 2 \, \Gamma \, R'_B / z)$, where 
$\Delta r_{BLR}$ is a measure of the geometrical thickness of the 
broad line region. Similar to the SSC spectrum, the RSy spectrum 
does not show a strong spectral break. It peaks around 
$\langle\epsilon\rangle_{RSy} \approx (B' / B_{cr}) \, D \, \Gamma^2 
\, \gamma_e^4 \approx \langle \epsilon \rangle_{SSC} \, \Gamma^2$. 

The energy density of the radiation field provided by IR emission
from circumnuclear dust is $u'_{IR} \sim 4 \, \Gamma^2 \, \xi_{IR}
\sigma T_{dust}^4 / c$, where $\xi_{IR}$ is the fraction of the flux 
from the central source (accretion disc) reprocessed by the warm dust
of temperature $T_{dust}$. Both $\xi_{IR}$ and $T_{dust}$ depend on 
the characteristic distance $r_{IR}$ of the dust from the central
source (Bla$\dot{\rm z}$ejowski et al. 2000). The resulting Compton
scattered radiation peaks around $\langle\epsilon\rangle_{ECIR}
\sim 3 \, k T_{dust} \, D \, \Gamma \, \gamma_e^2$.

\vskip 20 pt
\centerline{\bf 3. Spectral modeling of FSRQs and BL Lacs}
\bigskip 
\noindent
The leptonic jet models described in the previous section have
been used very successfully to model simultaneous broadband
spectra of several FSRQs, LBLs, and HBLs. As more detailed
spectral information has become available, the results of such 
broadband spectral modeling are now converging towards a rather 
consistent picture (e.g., Ghisellini et al. 1998, Kubo et al. 1998): 
The spectral sequence from HBLs to LBLs and on to FSRQs appears 
to be related to an increasing contribution of the external 
Comptonization mechanisms ECD and ECC to the $\gamma$-ray
spectrum. While most FSRQs are successfully modelled with external
Comptonization models (e. g., Dermer et al. 1997, Sambruna et al.
1997, Mukherjee et al. 1999, Hartman et al. 2001a), the broadband 
spectra of HBLs are consistent with pure SSC models (e. g., 
Mastichiadis \& Kirk 1997, Pian et al. 1998, Petry et al. 2000). 
BL~Lacertae, a LBL, appears to be intermediate between these 
two extremes, requiring an external Comptonization component 
to explain the EGRET spectrum (Madejski et al. 1999, B\"ottcher 
\& Bloom 2000). As an example, Fig. 3 shows a fit to the 
simultaneous broadband spectrum of 3C~279 during the 1991
June flare state (Hartman et al. 1996, Hartman et al. 2001a).

\midinsert
\epsfysize=6cm
\hskip 3cm \epsffile{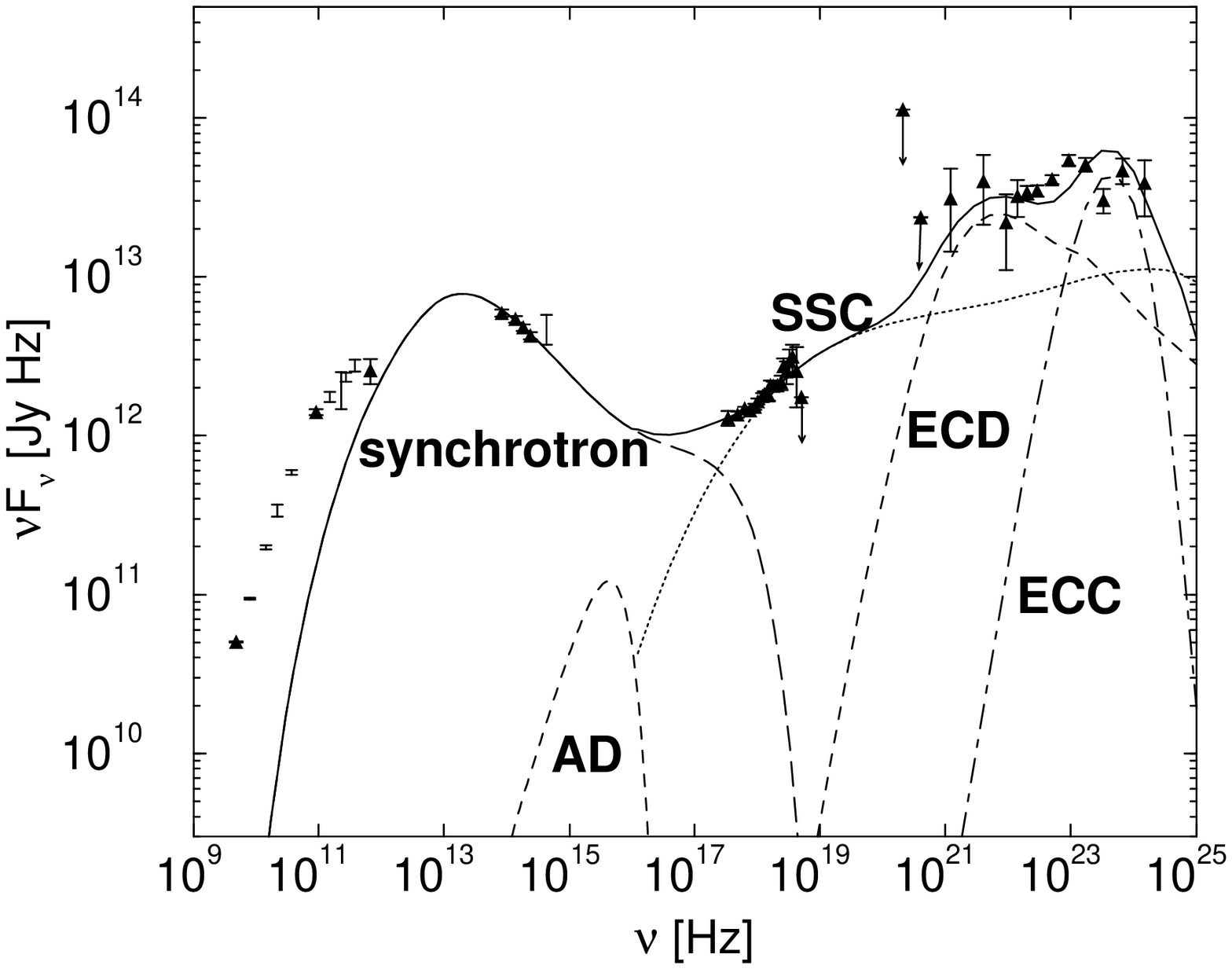}
\medskip
\noindent
{\bf Figure 3.} Fit to the simultaneous broadband spectrum of
3C~279 during the flare of 1991 June (Hartman et al. 1996).
For model parameters see Hartman et al. (2001a).
\endinsert

A physical interpretation of this sequence in the framework of a
unified jet model for blazars was given in Ghisellini et al. (1998).
Assume that the average energy of electrons, $\gamma_e$, is
determined by the balance of an energy-independent acceleration
rate $\dot\gamma_{acc}$ and radiative losses, $\dot\gamma_{rad}
\approx - (4/3) \, c \, \sigma_T \, (u' / m_e c^2) \, \gamma^2$,
where the target photon density $u'$ is the sum of the sources
intrinsic to the jet, $u'_B + u'_{sy}$ plus external photon
sources, $u'_{ECD} + u'_{ECC} + u'_{RSy}$. The average electron
energy will then be $\gamma_e \propto (\dot\gamma_{acc} / u')^{1/2}$.
If one assumes that the properties determining the acceleration
rate of relativistic electrons do not vary significantly between 
different blazar subclasses, then an increasing energy density
of the external radiation field will obviously lead to a
stronger radiation component due to external Comptonization,
but also to a decreasing average electron energy $\gamma_e$,
implying that the peak frequencies of both spectral components
are displaced towards lower frequencies.

\vskip 20 pt
\centerline{\bf 4. Modeling of spectral variability of blazars}
\bigskip 
\noindent
As described in the Introduction, HBLs exhibit rather consistent
spectral variability patterns during flaring episodes, which have
been very successfully reproduced by time-dependent model simulations 
in the framework of SSC-dominated jet models. In particular, the 
observed time lags between flares at different frequencies can be
related to the magnetic field $B$ and the Doppler boosting factor
$D$ through 
$$t(\nu_1) - t(\nu_2) \propto B^{-3/2} D^{-1/2} \left(\nu_1^{-1/2} 
- \nu_2^{-1/2} \right) \eqno(1)$$ 
(Takahashi et al. 1996). Also, the clockwise rotation of the spectral 
states of Mrk~421 and PKS~2155-304 has been modeled with a finite time 
scale of injection and/or acceleration of ultrarelativistic electrons 
along the jet --- where, phenomenologically, the flare results primarily
from an increasing high-energy cut-off $\gamma_2$ of the injected electron 
distribution ---, followed by radiative cooling, primarily due to 
synchrotron emission (e.g., Kirket al. 1998, Georganopoulos \& Marscher 
1998a, Kataoka et al. 2000, Kusunose et al. 2000, Li \& Kusunose 2000). 

The modeling of spectral variability described above refers to the
synchrotron component of HBLs which often peaks in the X-ray regime,
where --- at least during flares --- useful spectral information is 
achievable within a few ksec of integration time. However, at 
$\gamma$-ray energies, such information generally requires integration 
times longer than the expected synchrotron cooling time scale. Thus,
at present, such detailed short-term variability models cannot be
compared directly to $\gamma$-ray observations. However, a comparison
of averaged broadband spectra over longer time scales may still
provide useful model constraints. Comparing detailed spectral fits 
to weekly averaged broadband spectra of Mrk~501 over a period of 
6 months, Petry et al. (2000) have found that TeV and hard X-ray 
high states on intermediate timescales are consistent with a 
hardening of the electron spectrum (decreasing spectral index) 
and an increasing number density of high-energy electrons, while 
the value of $\gamma_2$ has only minor influence on the weekly 
averaged spectra. Fig. 4 shows the expected track of the 
time-averaged fluxes (on $\sim$ weekly time scales) in the
RXTE ASM and at $> 1.5$~TeV in the case of a change of only
one model parameter in a basic SSC model. The relatively large 
scatter of the observed correlation between the weekly averaged
X-ray and the TeV $\gamma$-ray emission (see Figs. 6 and 7 of 
Petry et al. 2000) indicates that the long-term spectral
variability of Mrk~501 might be associated with a change of more
than one of the basic model parameters between high and low X-ray
and $\gamma$-ray states. 

\midinsert
\epsfysize=6cm
\hskip 3cm \epsffile{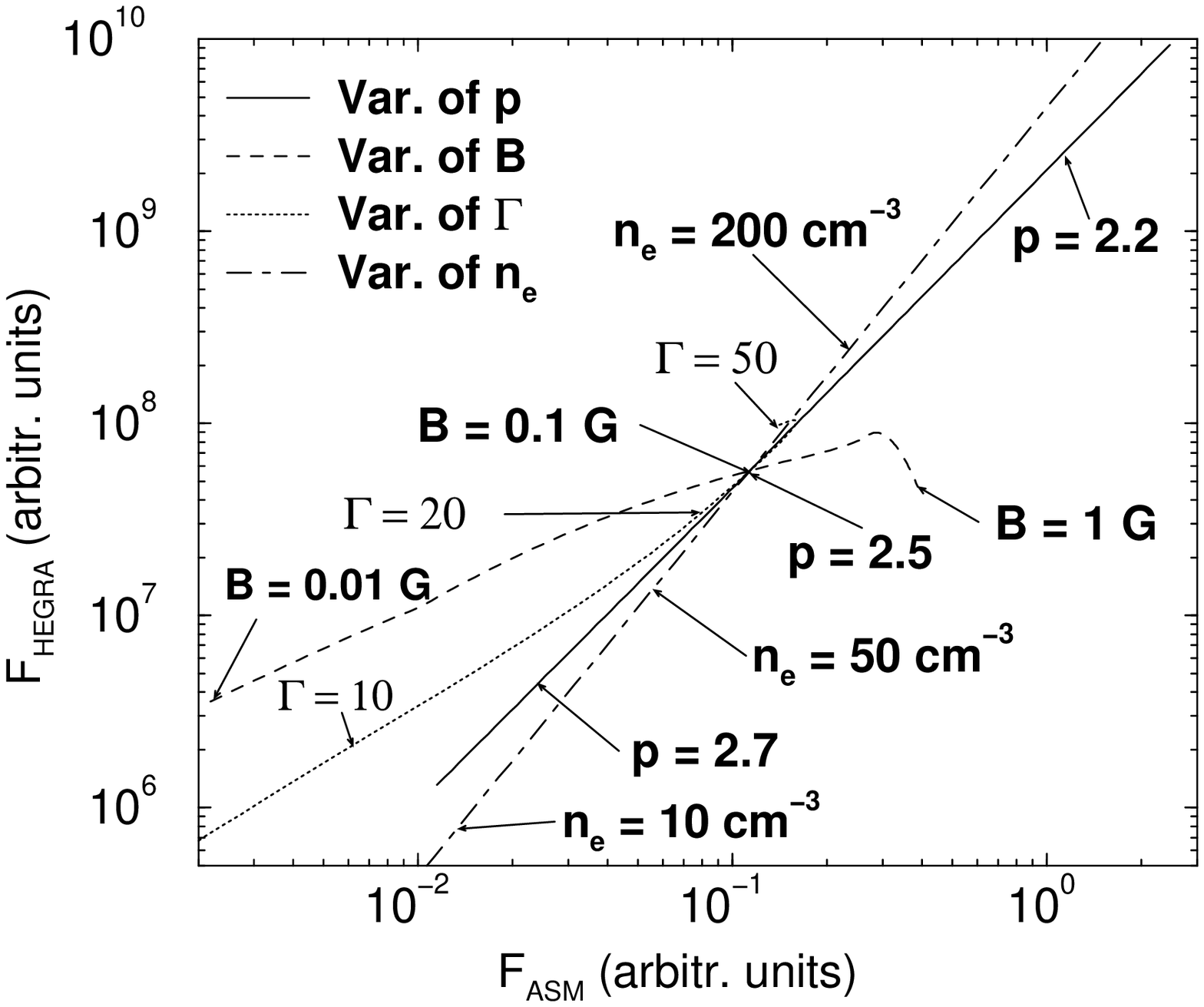}
\medskip
\noindent
{\bf Figure 4.} Expected correlation between the RXTE ASM and the
HEGRA $> 1.5$~TeV flux calculated from a time-averaged SSC model
similar to the analytic model of Tavecchio et al. (1998) --- from
Petry et al. (2000).
\endinsert

While the rapid variability of HBLs can be rather well modelled
and understood in the framework of SSC models, a similarly detailed
and meaningful analysis of the rapid variability of quasars is
until now hampered by (a) the lack of consistent variability
patterns among different objects and even within the same object,
and (b) the larger number of parameters entering the multi-component
high-energy spectral calculations required to provide acceptable
fits to FSRQ broadband spectra. In particular, in the framework
of leptonic jet models, rapid variability will be influenced by
(1) temporal variations of the seed photon fields for Comptonization
(see, e.g., B\"ottcher \& Dermer 1998 for a discussion of the
isolated effect of such variability), (2) time-dependent electron
acceleration and/or injection, and (3) time-dependent electron 
cooling. Given the limited time and frequency coverage available 
in simultaneous broadband observing campaigns to date, it is 
virtually impossible to disentangle the isolated effects of 
these causes of variability on the basis of currently available 
data.

A promising first step towards understanding the rapid variability of
quasars has recently been undertaken by Sikora et al. (2001), who have 
investigated a time-dependent shock-in-jet model with parameters
appropriate to reproduce FSRQ-like broadband spectra. The basic
idea of extracting physical information from observed variability
patterns in quasars is that within a spectral region dominated by 
the contribution from only one radiation mechanism, time-lag features 
similar to the ones seen in the X-ray -- UV -- optical spectra in
HBLs should result. From the fact that this is not consistent with
the correlated X-ray / $\gamma$-ray variability of 3C~279, Sikora
et al. (2001) conclude that the X-ray emission from this object
is dominated by a different mechanism --- most likely SSC --- than
the $\gamma$-rays, which might be dominated by Comptonization of
external radiation. This agrees very well with the spectral modeling
results of Hartman et al. (2001a).

\midinsert
\epsfysize=5.5cm
\epsffile{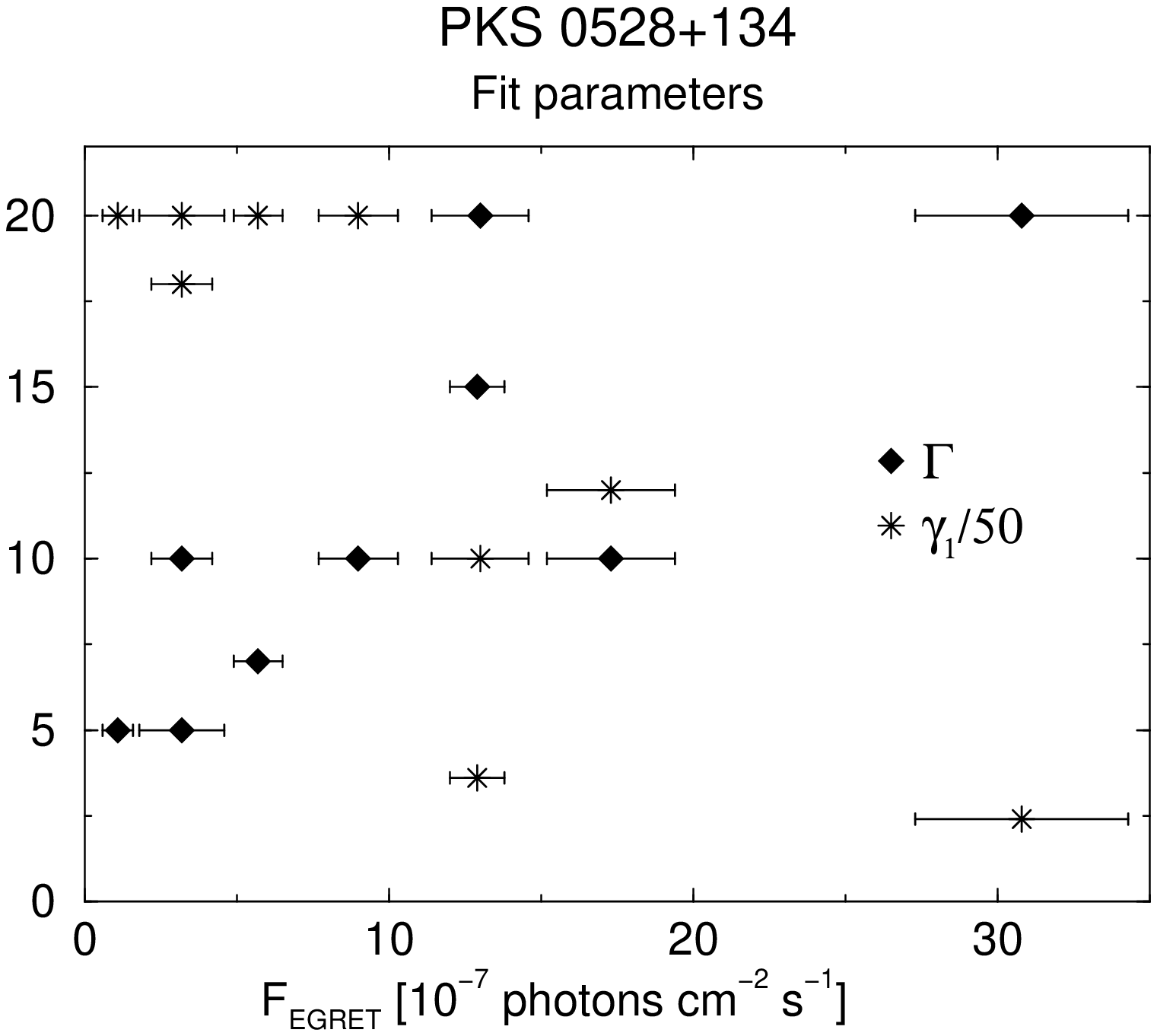}
\hskip 0.5cm
\epsfysize=5.5cm
\epsffile{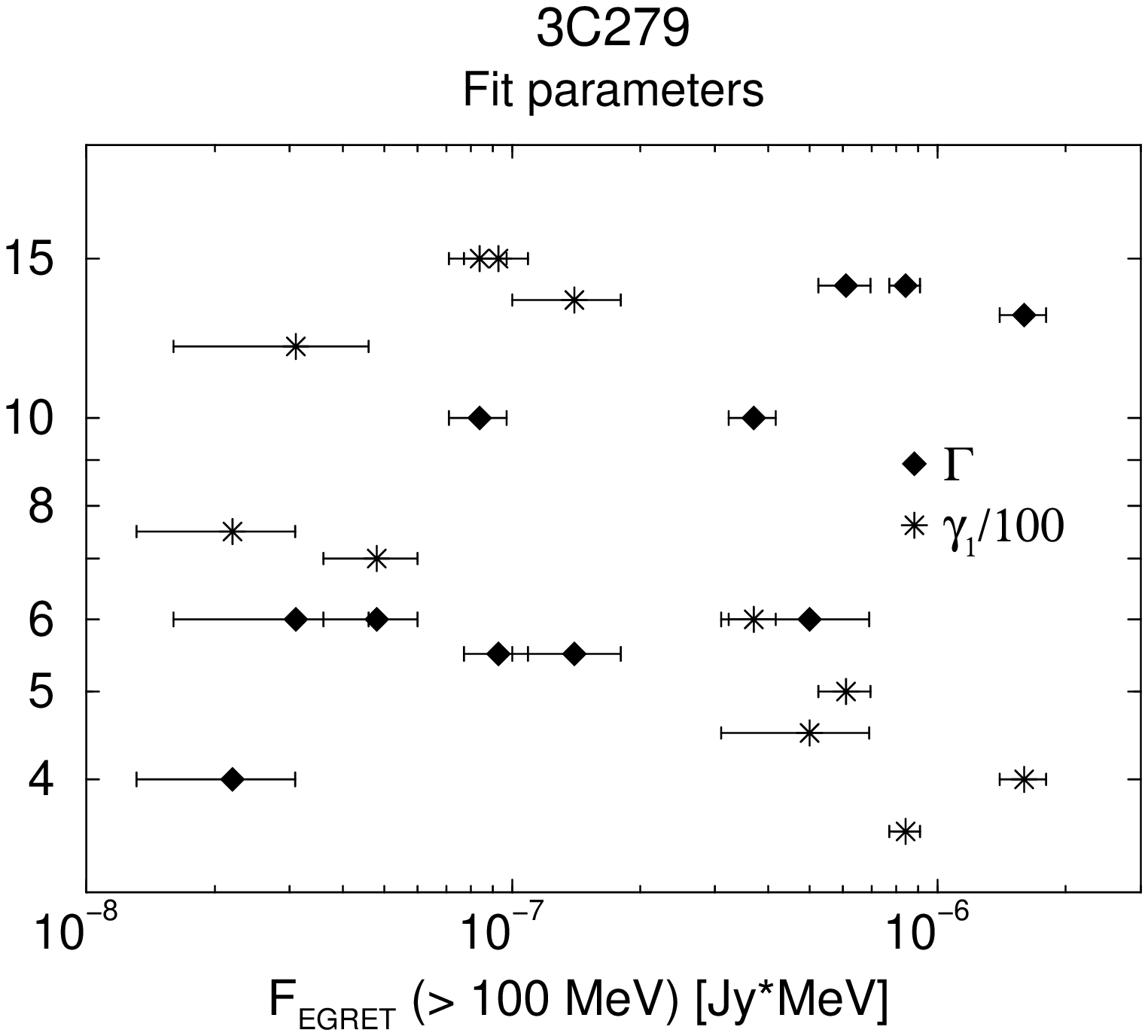}
\medskip
{\bf Figure 5.} Model fit parameters $\gamma_1$ (low-energy cut-off
of the electron distribution) and $\Gamma$ (bulk Lorentz factor) for
PKS~0528+134 and 3C~279 during different $\gamma$-ray intensity states
(Mukherjee et al. 1999, Hartman et al. 2001a).
\endinsert

In the case of quasars it is even more difficult to obtain time-dependent 
spectral information near the $\gamma$-ray peak, which is generally
located in the GeV regime, only accessible to satellite-borne instruments.
Even for the brightest quasar flares observed by EGRET, useful spectral
information could only be extracted on time scales of a few days. However,
during its $\sim 9$~year lifetime EGRET has observed several $\gamma$-ray
quasars multiple times in drastically different $\gamma$-ray intensity
states, so that a meaningful broadband variability study of those objects 
can be done by comparing simultaneous broadband spectra in different
EGRET observing periods. Such an analysis, accompanied with detailed
spectral modeling of each individual simultaneous broadband spectrum,
has been done for PKS~0528+134 (Mukherjee et al. 1999) and 3C~279
(Hartman et al. 2001a). In order to model the broadband spectra, as
many model parameters as possible have been held fixed between
different viewing periods. The long-term variability of those two
objects could be reasonably well reproduced with a higher bulk
Lorentz factor and a lower low-energy cut-off $\gamma_1$ of the
electron distribution during higher $\gamma$-ray states. Fig. 5
illustrates these trends. They bear a striking similarity to the 
trend between different blazar sub-classes in the sense that a
higher $\gamma$-ray luminosity indicates a stronger contribution
from Comptonization of external radiation, accompanied by a lower
value of $\gamma_1$. The analogy is mediated by the stronger
$\Gamma$-dependence of the external Compton radiation component 
compared to the SSC component (Dermer 1995), leading to an 
increasing contribution from external-radiation Comptonization 
with increasing $\Gamma$. 

However, it needs to be pointed out that due to the rather large
number of parameters necessary to achieve acceptable broadband
fits, other driving mechanisms for the long-term spectral variability
of those two quasars (for example, a change of the magnetic-field 
configuration combined with other plausible parameter changes,
which have not been investigated in Mukherjee et al. [1999] and
Hartman et al. [2001a]) can not be ruled out at this point. 

\vskip 20 pt
\centerline{\bf 5. Challenges for future observations}
\bigskip 
\noindent
In the previous sections, it has become clear that in particular
the blazar sub-classes of FSRQs and LBLs still pose many unsolved
questions as to the composition of their high-energy spectra and
the dominant acceleration and cooling mechanisms for relativistic
electrons in the jet. Two spectral regimes appear to be critical
to better understand those mechanisms:

(1) More continuous frequency and time coverage of quasar spectral
variability at optical -- UV -- soft X-ray frequencies would allow
the identification of acceleration- and cooling-related variability
patterns in the synchrotron component of FSRQs and LBLs. In particular,
on the basis of a comparison of optical/UV and soft X-ray variability 
one would be able to pin down the transition between the emission from
slow to fast cooling electrons and thus obtain accurate estimates of
the magnetic fields and Doppler factors of quasar and LBL jets as well
as additional size scale constraints. Dedicated monitoring observations 
with Chandra and XMM-Newton would be the ideal tool to achieve the
X-ray aspect of this challenge.

(2) In the most successful spectral quasar and LBL models, one expects
the transition between SSC-dominated and external-Compton-dominated
emission in the hard X-ray / soft $\gamma$-ray regime, in which it is
hard to achieve high-sensitivity observations with moderate spectral
resolution. However, this would be required in order to confirm the
predicted spectral hardening between the SSC and the external-Compton
components. This transition should also have very characteristic
imprints on the variability patterns, as pointed out by Sikora et
al. (2001). Unfortunately, it seems unlikely that the upcoming
INTEGRAL mission has sufficient sensitivity to allow spectral and
variability studies of quasars and LBLs in the $\sim 100$~keV -- 
10~MeV regime to place severe constraints on current models. Planned,
more sensitive future missions, such as EXIST and the ACT, may be 
required to achieve those goals.

\bigskip
\centerline{\bf References}
\bigskip
{\eightpoint\parindent=0pt\everypar={\hangindent=0.5 cm}
 Aharonian, F., 2000, New Astron., 5, 377.
 
 Arbeiter, C., Pohl, M., \& Schlickeiser, R., 2001, A\&A, submitted.
 
 Bednarek, W., 1998, A\&A, 342, 69.
 
 Blandford, R. D., \& Levinson, A., 1995, ApJ, 441, 79.
 
 Bla$\dot{\rm z}$ejowski, M., et al., 2000, ApJ, 545, 107.
 
 Bloom, S. D., \& Marscher, A. P., 1996, ApJ, 461, 657.

 B\"ottcher, M., 1999, ApJ, 515, L21.
 
 B\"ottcher, M., \& Bloom, S. D., 2000, AJ, 119, 469.
 
 B\"ottcher, M., \& Dermer, C. D., 1998, ApJ, 501, L51.
 
 B\"ottcher, M., Mause, H., \& Schlickeiser, R., 1997, A\&A, 324, 395.
 
 Bradbury, S. M., et al., 1997, A\&A, 320, L5.
 
 Catanese, M., et al., 1998, ApJ, 501, 616.
 
 Chadwick, P. M., et al., 1999, ApJ, 513, 161.
 
 Dermer, C. D., 1995, ApJ, 446, L63.
 
 Dermer, C. D., Schlickeiser, R., \& Mastichiadis, A., 1992, A\&A, 256, L27.
 
 Dermer, C. D., \& Schlickeiser, R., 1993, ApJ, 416, 458.
 
 Dermer, C. D., Sturner, S. J., \& Schlickeiser, R., 1997, ApJS, 109, 103.
 
 Edelson, R., et al., 1995, ApJ, 438, 120.
 
 Gaidos, J. A., et al., 1996, Nature, 383, 319.
 
 Georganopoulos, M., \& Marscher, A. P., 1998a, ApJ, 506, L11.
 
 Georganopoulos, M., \& Marscher, A. P., 1998b, ApJ, 506, 621.
 
 Ghisellini, G., \& Madau, P., 1996, MNRAS, 280, 67.
 
 Ghisellini, G., et al., 1998, MNRAS, 301, 451.
 
 Hartman, R. C., et al., 1996, ApJ, 461, 698.
 
 Hartman, R. C., et al., 1999, ApJS, 123, 79.
 
 Hartman, R. C., et al., 2001a, ApJ, 553, in press.
 
 Hartman, R. C., et al., 2001b, ApJ, in press.
 
 Kataoka, J., et al., 2000, ApJ, 528, 243.
 
 Kirk, J. G., Rieger, F. M., \& Mastichiadis, A., 1998, A\&A, 333, 452.
 
 Kubo, H., et al., 1998, ApJ, 504, 693.
 
 Kusunose, M., Takahara, F., \& Li, H., 2000, ApJ, 536, 299.
 
 Li, H., \& Kusunose, M., 2000, ApJ, 536, 729
 
 Madejski, G., et al., 1999, ApJ, 521, 145.
 
 Mannheim, K., 1993, A\&A, 269, 67.
 
 Maraschi, L., Ghisellini, G., \& Celotti, A., 1992, ApJ, 397, L5.
 
 Marscher, A. P., \& Gear, W. K., 1985, ApJ, 298, 114.
 
 Mastichiadis, A., \& Kirk, J. G., 1997, A\&A, 320, 19.
 
 M\"ucke, A., \& Protheroe, R. J., 2001, Astrop. Phys., 15, 121.
 
 Mukherjee, R., et al., 1997, ApJ, 490, 116.
 
 Mukherjee, R., et al., 1999, ApJ, 527, 132.
 
 Petry, D., et al., 1996, A\&A, 311, L13.
 
 Petry, D., et al., 2000, ApJ, 536, 742.
 
 Pian, E., et al., 1998, ApJ, ApJ, 492, L17.
 
 Punch, M., et al., 1992, Nature, 358, 477.
 
 Quinn, J., et al., 1996, ApJ, 456, L83.
 
 Rachen, J., 1999, AIP Conf. Proc., 515, 41.
 
 Sambruna, R., et al., 1997, ApJ, 474, 639.
 
 Schlickeiser, R., 1996, Space Sci. Rev., 75, 299.
 
 Sikora, M., Begelman, M. C., \& Rees, M. J., 1994, ApJ, 421, 153.
 
 Sikora, M., et al., 2001, ApJ, in press.
 
 Takahashi, T., et al., 1996, ApJ, 470, L89.
 
 Tavecchio, F., Maraschi, L., \& Ghisellini, G., 1998, ApJ, 509, 608.
 
 Urry, C. M., et al., 1997, ApJ, 486, 799.
 
 von Montigny, C., et al., 1995, ApJ, 440, 525.
 
 Wehrle, A., et al., 1998, ApJ, 497, 178.
 
}                                         
\end